\documentclass{aa}
\usepackage{graphicx,natbib,amsmath,amssymb,url}
\usepackage[T1]{fontenc}
\usepackage{mathptmx}
\usepackage[colorlinks=true,citecolor=blue,urlcolor=blue]{hyperref}
\bibpunct{(}{)}{;}{a}{}{,}
\newcommand{\kms}{\ifmmode{~{\rm km\,s}^{-1}}\else{~km~s$^{-1}$~}\fi}
\newcommand{\msun}{\ifmmode{{\rm M}_{\odot}}\else{${\rm M}_{\odot}$~}\fi}
\newcommand{\msunit}{\ifmmode{M_{\odot}}\else{$M_{\odot}$~}\fi}
\newcommand{\lsun}{\ifmmode{{\rm L}_{\odot}}\else{${\rm L}_{\odot}$}\fi}

\begin{document}

\title{Simulations of protostellar collapse using multigroup radiation hydrodynamics. I. The first collapse}
\titlerunning{Protostellar collapse using multigroup RHD. I.}

\author{Neil Vaytet$^{1}$, Edouard Audit$^{2,3}$, Gilles Chabrier$^{1,4}$, Beno\^{i}t Commer\c{c}on$^{5,6}$ and Jacques Masson$^{1}$}
\authorrunning{Vaytet et~al.}

\institute{$^{1}$ Ecole Normale Sup\'{e}rieure de Lyon, CRAL (UMR CNRS 5574), 69364 Lyon Cedex 07, France\\
           $^{2}$ Maison de la Simulation, USR 3441,  CEA - CNRS - INRIA - Universit\'{e} Paris-Sud - Universit\'{e} de Versailles, 91191 Gif-sur-Yvette, France\\
           $^{3}$ CEA/DSM/IRFU, Service d'Astrophysique, Laboratoire AIM, CNRS, Universit\'{e} Paris Diderot, 91191 Gif-sur-Yvette, France\\
           $^{4}$ School of Physics, University of Exeter, Exeter, EX4 4QL, UK\\
           $^{5}$ Max Planck Institute for Astronomy, K\"{o}nigstuhl 17, 69117, Heidelberg, Germany\\
           $^{6}$ Laboratoire de radioastronomie, UMR 8112 du CNRS, \'{E}cole normale sup\'{e}rieure et Observatoire de Paris, 24 rue Lhomond, 75231 Paris Cedex 05, France\\
          }

\date{Received / Accepted}

\offprints{neil.vaytet@ens-lyon.fr} 
   
\abstract{Radiative transfer plays a major role in the process of star formation, the details of which are still not entirely understood.}{Many previous simulations of gravitational collapse of a cold gas cloud followed by the formation of a protostellar core have used a grey treatment of radiative transfer coupled to the hydrodynamics. However, dust opacities which dominate circumstellar extinction show large variations as a function of frequency. In this paper, we used a frequency-dependent formalism for the radiative transfer in order to investigate the influence of the opacity variations on the properties of Larson's first core.}{We used a multigroup $M_{1}$ moment model for the radiative transfer in a 1D Lagrangean Godunov radiation hydrodynamics code to simulate the spherically symmetric collapse of a 1~\msun cold cloud core. Monochromatic dust opacities for five different temperature ranges were used to compute Planck and Rosseland means inside each frequency group.}{The results are very consistent with previous studies and only small differences were observed between the grey and multigroup simulations. For a same central density, the multigroup simulations tend to produce first cores with a slightly higher radius and central temperature. We also performed simulations of the collapse of a 10 and 0.1~\msun cloud, which showed that the properties of the first core (size, mass, entropy etc\ldots) are independent of the initial cloud mass, with again no major differences between grey and multigroup models.}{For Larson's first collapse, where temperatures remain below 2000 K, the vast majority of the radiation energy lies in the infrared regime and the system is optically thick. In this regime, the grey approximation does a good job reproducing the correct opacities, as long as there are no large opacity variations on scales much smaller than the width of the Planck function. The multigroup method is however expected to yield further more important differences in the later stages of the collapse when high energy (UV and X-ray) radiation is present and matter and radiation are strongly decoupled.}


\maketitle

\section{Introduction}

The process of star formation has been a subject of active research over the past few decades. It involves the gravitational collapse of a cold dense core inside a molecular cloud which heats up in its centre as the pressure and density increase from the compression, a problem which entails many physical processes (hydrodynamics, radiative transfer, magnetic fields, etc...) over a very large range of spatial scales (over 5 orders of magnitude) \citep{larson1969,stahler1980,masunaga1998}. The collapsing material is initially optically thin to the thermal emission from the cold gas and dust grains and all the energy gained from compressional heating is transported away by the escaping radiation, which causes the cloud to collapse isothermally in the initial stages of the formation of a protostar. When the optical depth of the cloud reaches unity, the radiation is absorbed by the system which starts heating up, taking the core collapse through its adiabatic phase. The strong compression forms an accretion shock at the border of the adiabatic core (also known as Larson's first core).

This radiative shock is believed to play a major role in the star formation process since it can either convert the kinetic energy of the material infalling onto the core to thermal energy which gets deposited inside the core or radiate this energy away, depending on the strength of the shock and the opacity of the gas. This largely determines key properties of the first core such as its size, mass and entropy level as well as the magnitude of the radiative flux at the core border \citep{winkler1980,commercon2011}. The presence of this radiative shock requires calculations to include the full radiation hydrodynamics (RHD) system of equations, which makes computations demanding. Three-dimensional (3D) RHD simulations have only just recently become possible with modern computers. The first 3D study of the collapse of an isolated spherical cloud using smoothed particle hydrodynamics and flux-limited diffusion (FLD) was carried out by \citet{whitehouse2006} \citep[see][for a description of the methodology]{whitehouse2004}. \citet{krumholz2007a} later used a grid-based AMR RHD code to simulate star formation in a turbulent molecular cloud, using FLD in the mixed frame formalism \citep{krumholz2007b}. More recently, with the continuous rise in computer power, several authors have been able to simulate star formation in increasingly large turbulent clouds \citep{bate2009,offner2009}, but such largescale 3D studies can never incorporate the same detailed physics as one-dimensional (1D) studies which are still very much essential to uncover and isolate all the crucial physical processes at work.

In this paper, we improve on the recent 1D calculations of \citet{commercon2011} by including frequency-dependent radiative transfer in our model as opposed to the more simple grey approximation which integrates the equations of radiative transfer over the entire frequency range. The motivation behind such an extension lies in the strong dependence of the circumstellar material opacities as a function of frequency. Indeed, large changes in opacities (several orders of magnitude!) from the radio to the X-ray part of the electromagnetic spectrum are ubiquitous to interstellar gas and dust (see for example \citealt{ossenkopf1994,li2001,draine2003,semenov2003}; and Fig.~\ref{fig:kappanu}). This implies that the cloud core and circumstellar envelope can be optically thick to some parts of the spectrum and optically thin to others at the same time. This has potentially important consequences on the evolution and properties of the protostellar core.

Of all the works cited above, \citet{masunaga1998} is the only one not to use the grey approximation. They implemented frequency-dependent radiative transfer in their 1D study by solving the transfer equation directly. They looked at exactly the same problem as the one discussed in this paper, and already addressed the astrophysical questions 15 years ago. However, their method cannot be realistically incorporated in future 3D simulations due to the overly complex nature of the full 7-dimensional radiative transfer equation. \citet{yorke2002} used a simpler multi-frequency FLD in 2D in the context of high mass ($> 30~\msun$) star formation through gravitational collapse, but once again, solving the radiative transfer at every frequency would be too computationally expensive in 3D. On the other hand, the multigroup approach, for which the frequency domain is divided into a finite number of bins or groups and the opacities are averaged within each group, enables us to limit the number of frequency bins. This proves to be a good compromise between reproducing the important aspects of frequency-dependent opacities and saving computational cost. Moreover, using limited numbers of frequency groups allows us to use a more accurate radiative transfer model than the FLD, such as the $M_{1}$ model (see below).

We first describe the multigroup radiative transfer model, the opacities and the numerical method used in the simulations. The results are then presented and compared to grey simulations highlighting the differences between the two schemes.

\section{The multigroup RHD collapse simulations}\label{sec:simulations}

\subsection{The equations of multigroup radiation hydrodynamics}\label{sec:rhd_equations}

The numerical method employed to solve the equations of RHD is the one described in \citet{vaytet2011}. We use the $M_{1}$ moment model for radiative transfer \citep{levermore1984,dubroca1999,gonzalez2007} in its multigroup formalism, where the equations of radiative transfer are integrated into a finite number of frequency groups and the opacities are averaged over the same frequency ranges. This allows us to account for frequency dependence of the absorption and emission coefficients. The system of multigroup RHD equations in the comoving frame (all the radiative quantities, including the frequencies, are expressed in the frame of reference moving with the fluid) is
\begin{alignat}{2}
\partial_t \rho             & + \nabla \cdot (\rho \mathbf{u})                                  & & =  0 \label{eq:cons_mass}\\
\partial_t(\rho \mathbf{u}) & + \nabla \cdot (\rho \mathbf{u} \otimes \mathbf{u} + p\mathbb{I}) & & = \sum_{g=1}^{Ng} (\sigma_{Fg}/c) \mathbf{F}_{g} - \rho \nabla \Phi \label{eq:cons_mom}\\
\partial_t e                & + \nabla \cdot (\mathbf{u}(e + p))                                & & = \sum_{g=1}^{Ng} \Big[ c\Big(\sigma_{Eg} E_{g} - \sigma_{Pg} \Theta_{g}(T)\Big)\notag\\
                          ~ &           ~~~~~~~~~~~~~~~~~~~~~~~~~~~~~~~~~~~   + ~ & & (\sigma_{Fg}/c) \mathbf{u} \cdot \mathbf{F}_{g} \Big] - \rho \mathbf{u} \cdot \nabla \Phi\label{eq:cons_ener}\\
\partial_{t} E_{g}          & + \nabla \cdot \mathbf{F}_{g} + \nabla \cdot (\mathbf{u} E_{g}) +  \mathbb{P}_{g} & & : \nabla \mathbf{u} \notag\\
- \nabla \mathbf{u}         & : \displaystyle \int_{\nu_{g-1/2}}^{\nu_{g+1/2}}\partial_{\nu}(\nu \mathbb{P}_{\nu}) d\nu  & & = c \big( \sigma_{Pg} \Theta_{g}(T) - \sigma_{Eg} E_{g} \big) \label{eq:cons_Er}\\
\partial_{t} \mathbf{F}_{g} & + c^{2} \nabla \cdot \mathbb{P}_{g} + \nabla \cdot (\mathbf{u} \otimes \mathbf{F}_{g}) & & + \mathbf{F}_{g} \cdot \nabla \mathbf{u} \notag\\
- \nabla \mathbf{u}         &  : \displaystyle \int_{\nu_{g-1/2}}^{\nu_{g+1/2}}\partial_{\nu}(\nu \mathbb{Q}_{\nu}) d\nu  & & = - \sigma_{Fg} c \mathbf{F}_{g} \label{eq:cons_Fr}
\end{alignat}
where $\rho$, $\mathbf{u}$, $p$ and $e$ are the gas density, velocity, pressure and total energy, respectively, and $\Phi$ is the gravitational potential. We also define
\begin{equation}\label{eq:groupvar}
X_{g} = \int_{\nu_{g-1/2}}^{\nu_{g+1/2}} X_{\nu} d\nu
\end{equation}
where $X_{g} = E_{g}$, $\mathbf{F}_{g}$, $\mathbb{P}_{g}$ which represent the radiative energy, flux and  pressure inside each group $g$ which holds frequencies between $\nu_{g-1/2}$ and $\nu_{g+1/2}$. $\mathbb{Q}$ is the third moment of the radiation specific intensity (representing a heat flux), $N_{g}$ is the total number of groups and $\Theta_{g}(T)$ is the energy of the photons having a Planck distribution at temperature $T$ inside a given group. The absorption coefficients  $\sigma_{Pg}$, $\sigma_{Eg}$ and $\sigma_{Fg}$ are the means of $\sigma_{\nu}$ inside a given group weighted by the Planck function, the radiative energy and the radiative flux, respectively. The opacity $\kappa_{\nu}$ of the gas in defined as $\sigma_{\nu} = \rho\kappa_{\nu}$. As in \citet{mihalas1984}, the notation $\mathbb{P}:\nabla\mathbf{u}$ represents the tensorial product $P_{ij}\partial^{i}u^{j}$.

\subsection{The opacities}\label{sec:opacities}

\begin{figure}[!ht]
\centering
\includegraphics[scale=0.38]{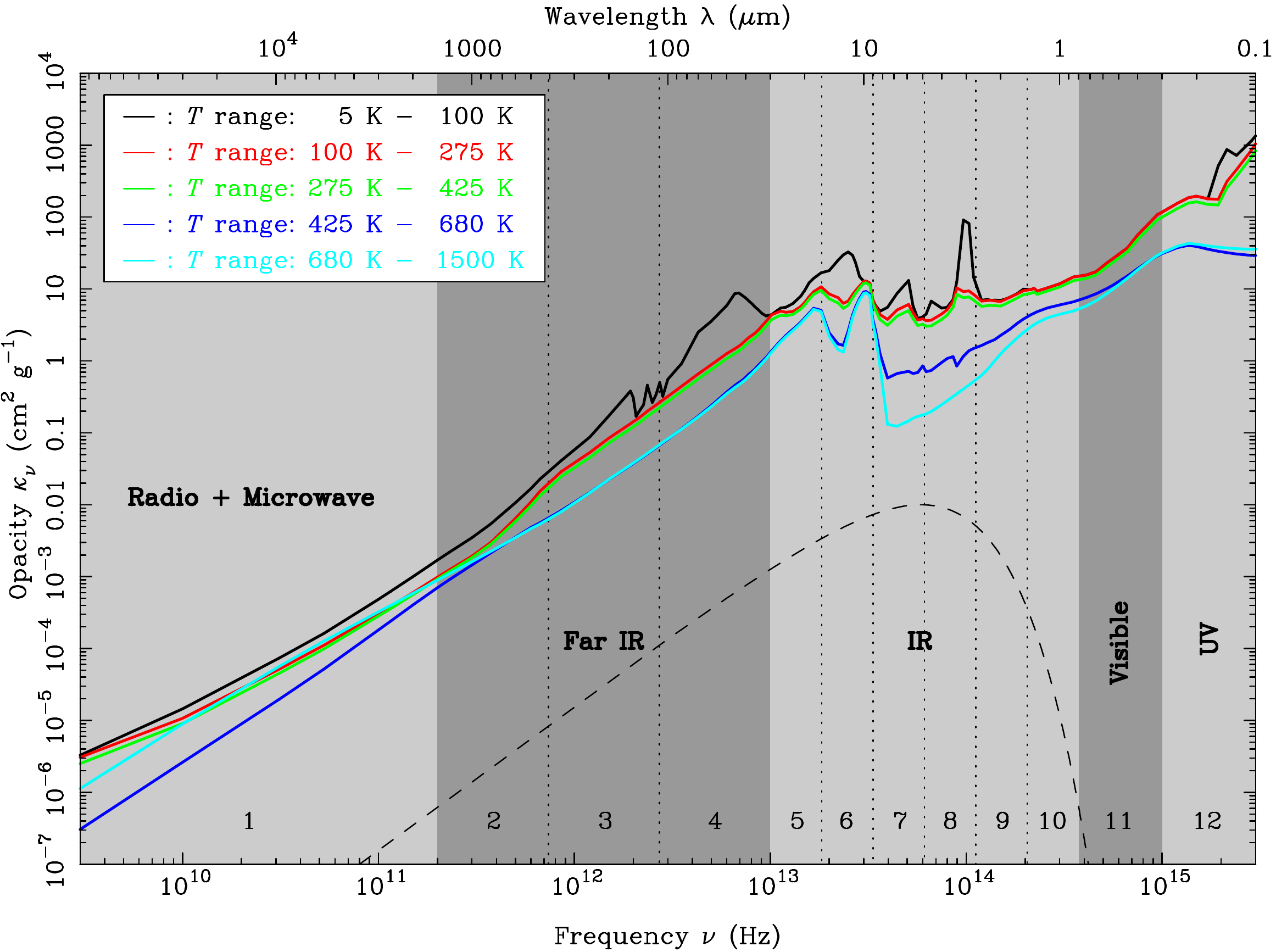}
\caption[Spectral opacities]{Spectral opacities for homogeneous spherical dust grains taken from \citet{semenov2003} for five different temperature ranges. The grey colour bands represent the five different frequency groups that were used in the multigroup simulation. The dotted vertical lines indicate the further splitting of groups 2 and 3 in the case of the 12-group simulation. The dashed line represents the Planck function for a black body with a typical temperature of 1000 K normalised to arbitrary units. The numbers in the lower part of the figure identify the different groups used in the 12-group simulation (see section~\ref{sec:results}).}
\label{fig:kappanu}
\end{figure}

At low temperatures (below 1500 K), the opacities of the interstellar gas are dominated by the one percent dust grains and the atomic and molecular gas opacities can be neglected. We used the monochromatic opacities from \citet{semenov2003}\footnote{\scriptsize\url{ http://www.mpia.de/homes/henning/Dust_opacities/Opacities/opacities.html}} who provide dust opacities for various types of grains in five different temperature ranges (we assume that the dust opacities are independent of gas density and that the dust is in thermal equilibrium with the gas; see \citealt{galli2002} for example). The spectral opacities for homogeneous spherical dust grains and normal iron content in the silicates ($\text{Fe}/(\text{Fe} + \text{Mg})=0.3$) are shown in Fig.~\ref{fig:kappanu}; the different line colours correspond to the different temperature ranges explicated in the legend. The grey-shaded areas represent the five frequency groups used in the multigroup simulation: radio + microwave, far IR, IR, visible and UV (see Table~\ref{tab:group-freqs}).

\begin{figure*}[!ht]
\centering
\includegraphics[scale=0.5]{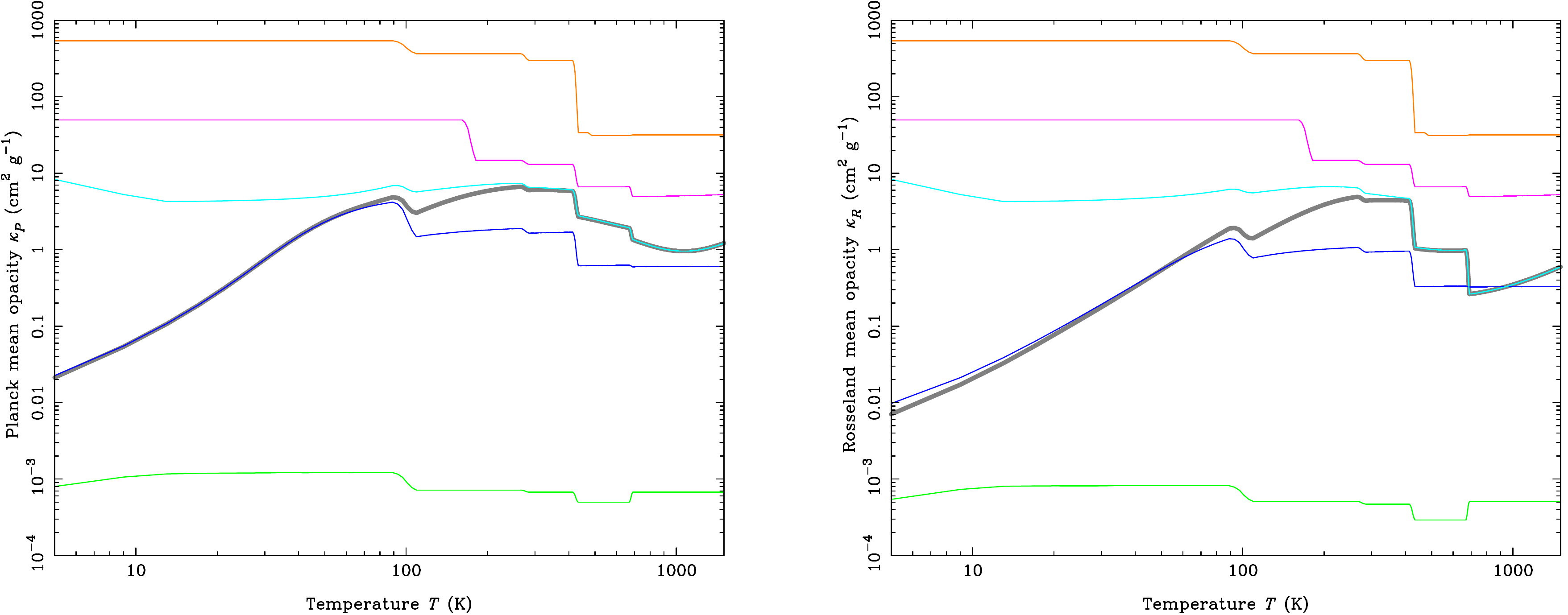}
\caption[Radial profiles of the first core]{Planck (left) and Rosseland (right) mean opacities as a function of temperature for the grey (thick grey) and multigroup (colours) simulations. The sharp discontinuities seen in the curves are due to transitions from one temperature range to the other, corresponding to the destruction of ice, silicates, etc\ldots}
\label{fig:kappaT}
\end{figure*}

\begin{table}
\centering
\caption[Group frequencies]{Lower and upper boundary frequencies for the five groups used in the multigroup simulation.}
\begin{tabular}{cccc}
\hline
\hline
Group  & Spectral          & Lower                 & Upper\\
number & region            & frequency (Hz)        & frequency (Hz)\\
\hline
1      & Radio + microwave & $0.00$                & $2.00 \times 10^{11}$\\
2      & Far IR            & $2.00 \times 10^{11}$ & $1.00 \times 10^{13}$\\
3      & IR                & $1.00 \times 10^{13}$ & $3.75 \times 10^{14}$\\
4      & Visible           & $3.75 \times 10^{14}$ & $1.00 \times 10^{15}$\\
5      & UV                & $1.00 \times 10^{15}$ & $3.00 \times 10^{15}$\\
\hline
\end{tabular}
\label{tab:group-freqs}
\end{table}

Planck ($\kappa_{P}$) and Rosseland ($\kappa_{R}$) mean opacities were calculated over the entire frequency range for the grey simulation and inside each frequency group for the multigroup simulation. We show in Fig.~\ref{fig:kappaT} the mean opacities which were tabulated for temperatures between 5 K and 1500 K using $\sim10$ K wide linear transition between the five different temperature regimes which correspond to the destruction of ice, silicates, etc\ldots \citep[see][]{semenov2003}. In the RHD equations (\ref{eq:cons_mom})--(\ref{eq:cons_Fr}), common practise is to set $\sigma_{Eg} = \sigma_{Pg}$ and $\sigma_{Fg} = \sigma_{Rg}$, and this is also what we have used in the present work \citep[see][for example]{larsen1994,gonzalez2007,offner2009}. However, we wish to point out that the inaccuracies which arise from this approximation are reduced as the number of frequency groups used increases, since in the limit of infinite frequency resolution, all of these quantities simply reduce to $\sigma_{\nu}$. This approximation is thus less crude in a multigroup model than in a grey model.

\subsection{Initial and boundary conditions}\label{sec:init_cond}

The initial setup for the dense core collapse is identical to \citet{commercon2011} and \citet{masunaga1998}. A uniform density sphere of mass $M_{0} = 1~\msun$, temperature $T_{0} = 10$ K ($c_{s0} = 0.187 \kms$) and radius $R_{0} = 10^{4}$ AU collapses under its own gravity. The ratio of thermal to gravitational energy in the cold gas cloud is
\begin{equation}
\alpha = \frac{5R_{0}k_{B}T_{0}}{2GM_{0}\mu m_{H}} = 0.98
\end{equation}
and its free-fall time is $t_{ff} \sim 0.177$ Myr. The radiation temperature is in equilibrium with the gas temperature (the energy of a black body with $T = 10$ K is divided among the frequency groups according to the Planck distribution) and the radiative flux is set to zero everywhere. The boundary conditions are reflexive at the centre of the grid ($r = 0$) and have imposed values equal to the initial conditions at the outer edge of the sphere. We used an ideal gas equation of state with the ratio of specific heats $\gamma = 5/3$ and the mean molecular weight $\mu = 2.375$ for molecular hydrogen and a Helium concentration of 0.27.

We also performed simulations of the collapse of 10 and 0.1~\msun clouds (see section \ref{sec:0.1-10msun}), where the setup is almost identical to the 1~\msun case. The thermal to gravitational energy ratio $\alpha$ was kept the same and the parent cloud radii were now $R_{0} = 10^{5}$ and $10^{3}$ AU, respectively. The free-fall times were then 10 times longer for the 10~\msun case and 10 times shorter for the 0.1~\msun case.

\subsection{Numerical method}\label{sec:num_method}

The numerical method is the spherically symmetric version of the one described in \citet{vaytet2011}; a 1D Lagrangean second order Godunov RHD code. The hydrodynamics equations are solved using a classical MUSCL-Hancock scheme while the radiative transfer equations are solved using a HLLC solver with an asymptotic preserving correction \citep{berthon2007}. The RHD equations are integrated with a scheme implicit in time, using a standard Raphson-Newton iterative method and the \textsc{lapack} library for the Jacobian matrix inversion. The radial (spherically symmetric) grid comprises 2000 cells and is logarithmically regular. The regridding scheme of \citet{dorfi1987} was also used (decoupled from the hydrodynamics) to maintain good resolution at the accretion shock.

\section{Results}\label{sec:results}

\subsection{The 1 \msunit case}\label{sec:1msun}

\begin{figure*}[!ht]
\centering
\includegraphics[scale=0.59]{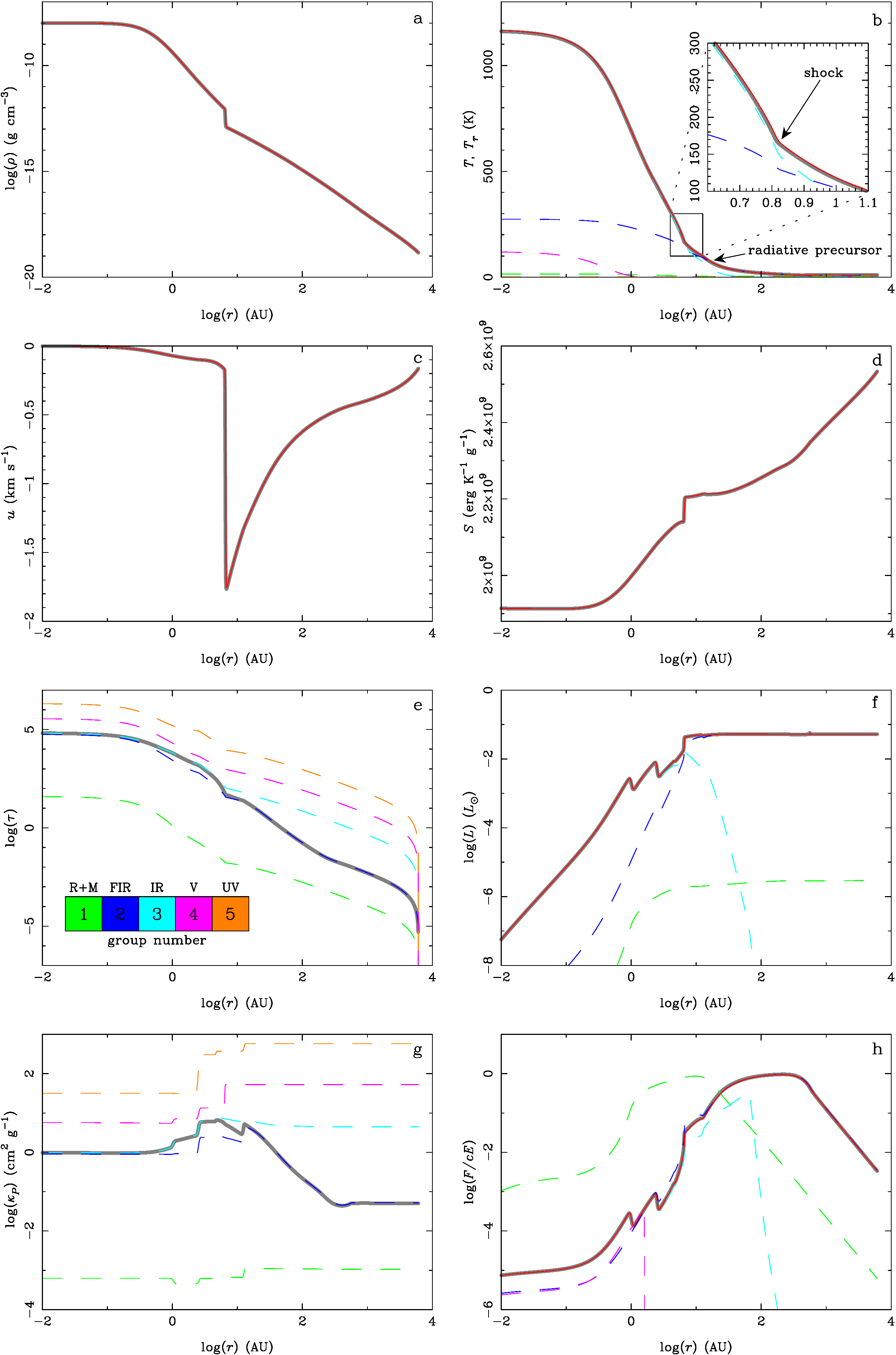}
\caption[First core radial profiles for 1 \msun cloud with 5 groups]{Radial profiles of various properties during the collapse of a $1~\msun$ dense clump at a core central density $\rho_{c} = 10^{-8}~\text{g~cm}^{-3}$, for the grey (grey thick solid line) and multigroup (colours) models. For the multigroup run, the gas and total radiative (summed over all groups) quantities are plotted in red, while the other colours represent the individual groups; see legend in (e). From top left to bottom right: (a) density, (b) gas (solid) and radiation (dashed) temperature, (c) velocity, (d) entropy, (e) optical depth, (f) luminosity, (g) Planck average opacity and (h) reduced radiative flux.}
\label{fig:collapse}
\end{figure*}

A grey and a multigroup simulations of the collapse of a dense core were performed. In the grey run, the radiative quantities were integrated over the entire frequency range (0 to $3 \times 10^{15}$ Hz) while for the multigroup run, five frequency groups were used (see Table~\ref{tab:group-freqs}). The grey run should be more or less equivalent to the simulation in \citet{commercon2011}, as a very similar method was used (the opacities and the computational grid differ slightly and the solver for the radiative transfer is different). The simulations were run until the central density reached $\rho_{c} = 10^{-8}~\text{g~cm}^{-3}$. Figure~\ref{fig:collapse} shows the radial profiles of the density, temperature, velocity, entropy, optical depth, luminosity, radiative flux and mass for the grey (grey thick solid line) and multigroup simulations (colours).

We first note that the results for the grey simulation are again very consistent with the results of \citet{masunaga1998}, as was noted in \citet{commercon2011}. An adiabatic core has formed at the centre of the grid with a shock clearly visible at its border, at a radius of $\sim 7$ AU (see Figs.~\ref{fig:collapse}a and \ref{fig:collapse}c for instance). From the temperature plot (b), we see that this is a supercritical radiative shock; the pre- and post-shock temperatures are identical and the radiative precursor is visible in between 7 and 100 AU. The properties of the first core are listed in Table~\ref{tab:core-props}; these are the first core radius ($R_{\text{fc}}$), the mass of the first core ($M_{\text{fc}}$), the mass accretion rate at the accretion shock
\begin{equation}\label{eq:mdot}
\dot{M} = 4\pi r^{2} \rho u ~~~,
\end{equation}
the accretion luminosity
\begin{equation}\label{eq:lacc}
L_{\text{acc}} = \frac{GM_{\text{fc}}\dot{M}}{R_{\text{fc}}} ~~~,
\end{equation}
the total luminosity ($L_{\text{tot}}$), the gas temperature at the centre of the core ($T_{\text{c}}$), the gas temperature at the accretion shock $T_{\text{fc}}$ and the entropy at the centre of the core
\begin{equation}\label{eq:sc}
S_{\text{c}} = \frac{k_{B}}{\mu m_{\text{H}} (\gamma - 1)} \ln \left( \frac{P}{\rho^{\gamma}} \right) ~~~.
\end{equation}
They are very close to that listed in \citet{commercon2011}, which is once again expected since we used almost the same computational method.

There are important variations in radiative flux (h) (and hence luminosity) inside the first core, which occur where there are large jumps in dust opacity (g) when changing from one temperature range to the next (destruction of ice, pyroxene and other dust grain constituents, see \citealt{semenov2003}), as the variation in temperature is important in this part of the system. The jump in luminosity at the position of the shock is due to the accretion luminosity which is added to the internal luminosity of the first core to constitute the total luminosity. The power radiated per unit area is
\begin{equation}
\frac{L}{A} = \frac{c}{4} f \int_{0}^{\infty} E_{\nu} d\nu
\end{equation}
and therefore the radiative luminosity just inside of the shock is $L_{\mathrm{rad}} = \pi R_{\mathrm{fc}}^{2} f E_{g} c = 1.7 \times 10^{-2}~\lsun$ for $A = 4\pi R_{\mathrm{fc}}^{2}$ in the case of one frequency group (we recall that $E_{g} = a_{R}T_{r}^{4}$ in this case). The accretion luminosity is $L_{\mathrm{acc}} = 2.16 \times 10^{-2}~\lsun$ (see Table~\ref{tab:core-props}) and the sum of the two approximately corresponds to the total radiative luminosity observed just outside of the shock $L_{\mathrm{tot}} = 4.2 \times 10^{-2}~\lsun \simeq L_{\mathrm{rad}} + L_{\mathrm{acc}}$. The missing $3 \times 10^{-3}~\lsun$ comes from the fact that the optical depth per cell falls abruptly at the shock because of the drop in gas density, and more radiation thus escapes from the core, causing a mini-burst in luminosity.

\begin{figure*}[!ht]
\centering
\includegraphics[scale=0.59]{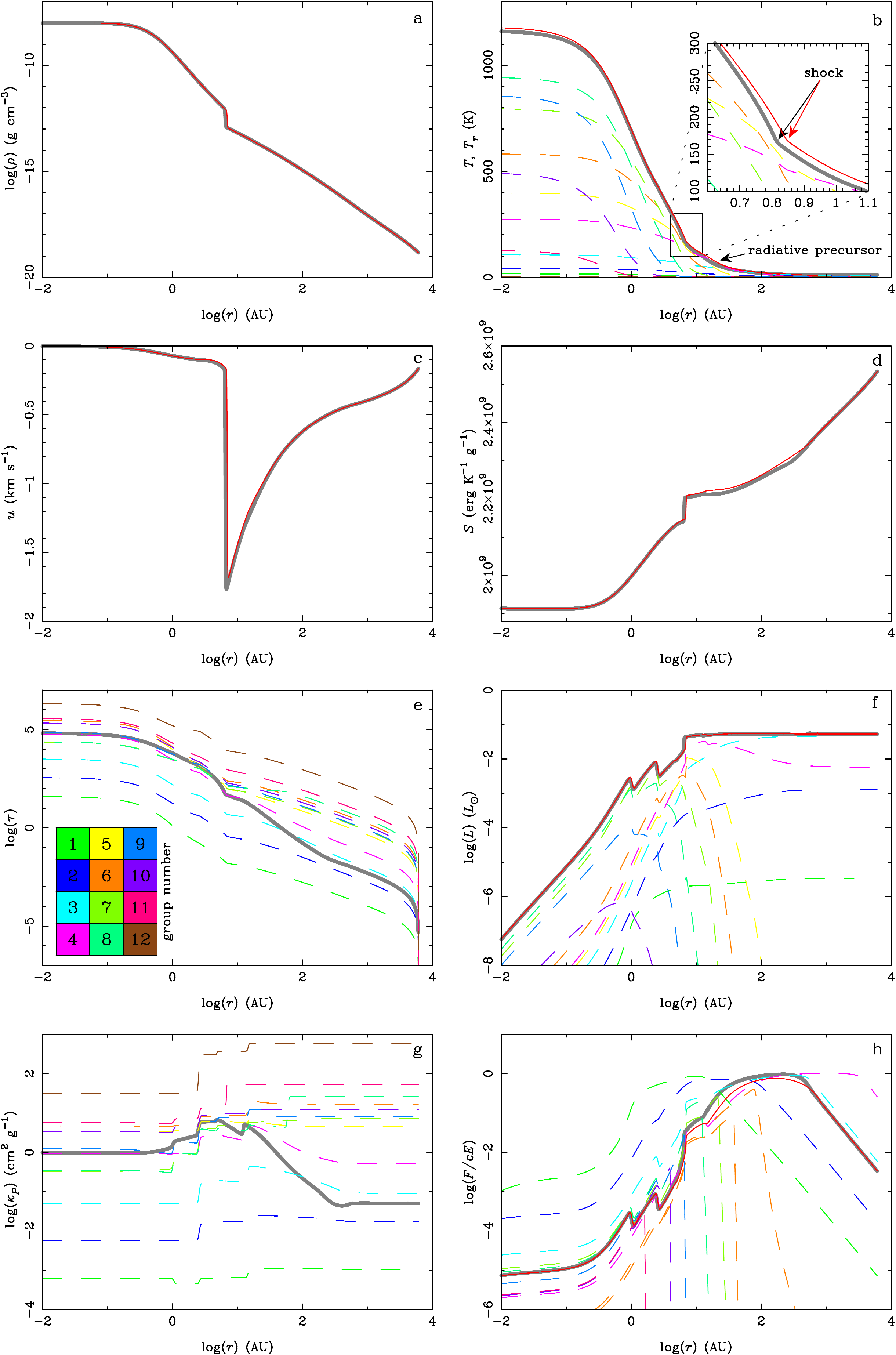}
\caption[First core radial profiles for 1 \msun cloud with 12 groups]{Same as Fig.~\ref{fig:collapse} but for the 12 group simulation.}
\label{fig:collapse12}
\end{figure*}

\begin{table*}
\centering
\caption[First core properties]{Summary of the first core properties when $\rho_{c} = 10^{-8}~\text{g~cm}^{-3}$ for the 0.1 \msun, 1 \msun and 10 \msun cases. The first column indicates the total mass of the parent gas cloud, the second specifies the number of frequency groups used in each run and the one before last indicates the upstream mach number of the flow at the accretion shock. The final column lists the time at which the density at the centre of the grid reached $\rho_{c}$ (the time is independent of the number of groups used). The other quantities are explicited in the text.}
\begin{tabular}{cccccccccccc}
\hline
\hline
Mass of & \# of & $R_{\text{fc}}$ & $M_{\text{fc}}$ & $\dot{M}$    & $L_{\text{acc}}$ & $L_{\text{tot}}$ & $T_{\text{c}}$ & $T_{\text{fc}}$ & $S_{\text{c}}$ & Mach & Time\\
cloud   & groups    & (AU)            & ($\msun$)       & ($\msun/\text{yr}$) & \lsun            & \lsun            & (K)            & (K)             & (erg/K/g) & number & (Myrs)\\
\hline
~          &  1 & 6.86 & $1.87 \times 10^{-2}$ & $2.50 \times 10^{-5}$ & $9.90 \times 10^{-3}$ & $2.51 \times 10^{-2}$ & 999 & 111 & $1.91 \times 10^{9}$ & 2.09 & ~\\
0.1 \msun  &  5 & 6.96 & $1.88 \times 10^{-2}$ & $2.48 \times 10^{-5}$ & $9.73 \times 10^{-3}$ & $2.49 \times 10^{-2}$ & 1002 & 113 & $1.91 \times 10^{9}$ & 2.07 & 0.020\\
~          & 12 & 7.54 & $1.92 \times 10^{-2}$ & $2.36 \times 10^{-5}$ & $8.78 \times 10^{-3}$ & $2.50 \times 10^{-2}$ & 1011 & 115 & $1.91 \times 10^{9}$ & 1.92 & ~\\
\hline
~          &  1 & 6.54 & $2.19 \times 10^{-2}$ & $4.44 \times 10^{-5}$ & $2.16 \times 10^{-2}$ & $4.21 \times 10^{-2}$ & 1164 & 167 & $1.91 \times 10^{9}$ & 1.82 & ~\\
1 \msun    &  5 & 6.64 & $2.20 \times 10^{-2}$ & $4.42 \times 10^{-5}$ & $2.14 \times 10^{-2}$ & $4.21 \times 10^{-2}$ & 1167 & 167 & $1.91 \times 10^{9}$ & 1.81 & 0.194\\
~          & 12 & 6.99 & $2.25 \times 10^{-2}$ & $4.29 \times 10^{-5}$ & $2.01 \times 10^{-2}$ & $4.23 \times 10^{-2}$ & 1180 & 170 & $1.91 \times 10^{9}$ & 1.72 & ~\\
\hline
~          &  1 & 6.42 & $2.36 \times 10^{-2}$ & $5.71 \times 10^{-5}$ & $3.06 \times 10^{-2}$ & $5.07 \times 10^{-2}$ & 1245 & 203 & $1.92 \times 10^{9}$ & 1.69 & ~\\
10 \msun   &  5 & 6.52 & $2.37 \times 10^{-2}$ & $5.71 \times 10^{-5}$ & $3.03 \times 10^{-2}$ & $5.10 \times 10^{-2}$ & 1249 & 202 & $1.92 \times 10^{9}$ & 1.68 & 1.930\\
~          & 12 & 6.76 & $2.42 \times 10^{-2}$ & $5.57 \times 10^{-5}$ & $2.91 \times 10^{-2}$ & $5.17 \times 10^{-2}$ & 1264 & 206 & $1.92 \times 10^{9}$ & 1.62 & ~\\
\hline
\end{tabular}
\label{tab:core-props}
\end{table*}

Secondly, and much more importantly, we note that there are virtually no differences between the grey and multigroup simulations. All the profiles are identical to their grey counterparts. We see that the IR frequency group (3) dominates inside the first core (radiative temperature and flux) while the far IR group (2) takes over in the outer envelope. One also notes that the grey Planck mean opacity (grey thick solid line in g) is driven towards the values in these groups by the weighting of the Planck function for the temperatures of the core ($200 - 1200$ K) and the outer envelope ($10 - 200$ K). As listed in Table~\ref{tab:core-props}, the first core radius and central temperature are very slightly larger in the multigroup simulation but these differences are of the order of one percent, hence insignificant for the global properties of the cores.

There is no jump in gas temperature at the accretion shock in both the grey and multigroup simulations, illustrating that the radiative shock is supercritical in both cases. This and the upstream Mach number of the shock ($M\sim2$) implies that the vast majority of the kinetic energy from the infalling material is radiated away at the shock \citep{commercon2011}.

In light of these small differences between the grey and 5-group simulation, we ran a third simulation, this time using 12 frequency groups. Since the groups 2 and 3 hold the majority of the radiative energy (and show large opacity variations), they were split further while groups 1, 4 and 5 were kept unchanged. Group 2 was split into three logarithmically equal parts and group 3 into 6 parts, as illustrated by the dotted vertical lines in Fig.~\ref{fig:kappanu}. The results for the 12-group simulation are compared to the grey run in Fig.~\ref{fig:collapse12} and the first core properties are listed in Table~\ref{tab:core-props}.

There are again no major differences between the 12-group and the two previous simulations. The first core radius is slightly larger in the 12-group case (difference of 10\% compared to the grey run) and the central temperature is also marginally higher, but the other first core properties are very similar to the 5-group simulation (the time needed to reach $\rho_{c}$ was also 0.194 Myr; identical to the 5-group run). Panels (b), (f) and (h) illustrate well the sharing of the radiative energy and flux among the different frequency groups. Groups $5-10$ hold the majority of the radiative energy inside the core, where for the most part, temperatures exceed 600 K. Figure~\ref{fig:collapse12}g shows that groups 5, 6 and 10 have a significantly higher opacity than the grey opacity, and that the opacities groups 7 and 8 are lower. This is also illustrated by Fig.~\ref{fig:kappanu} where groups 5 and 6 correspond to peaks in spectral opacity (light blue curve) while 7 and 8 are in the low-opacity region immediately following. The spectral opacities increase again in groups 9 and 10. This means that inside the core, there is a combination of high and low opacities, and that the grey opacity lies somewhere in the middle. This explains why the central temperatures differ between the grey and the 12-group case, and in this particular instance, the higher opacities in groups 5, 6 and 10 `win' over the lower opacities in 7 and 8, preventing a little more radiation from escaping and causing the core to be somewhat hotter. The heating of the core works against the gravitational contraction, allowing the core to have a larger radius. 

Overall, however, the variations in dust opacity as a function of frequency are not sharp enough to induce a large impact on the results, compared to the grey simulation which weighs the Planck opacity using the Planck function (and its derivative with respect to temperature in the case of the Rosseland opacity) over the entire frequency range and does a good job of reproducing the variation of the opacity as a function of gas temperature. Important differences between grey and multigroup simulations would appear if large variations in opacities (several orders of magnitude) occured on scales much smaller that the typical width of the Planck function.

\subsection{The 0.1 \msunit  and 10 \msunit cases}\label{sec:0.1-10msun}

\begin{figure*}[!ht]
\centering
\includegraphics[scale=0.60]{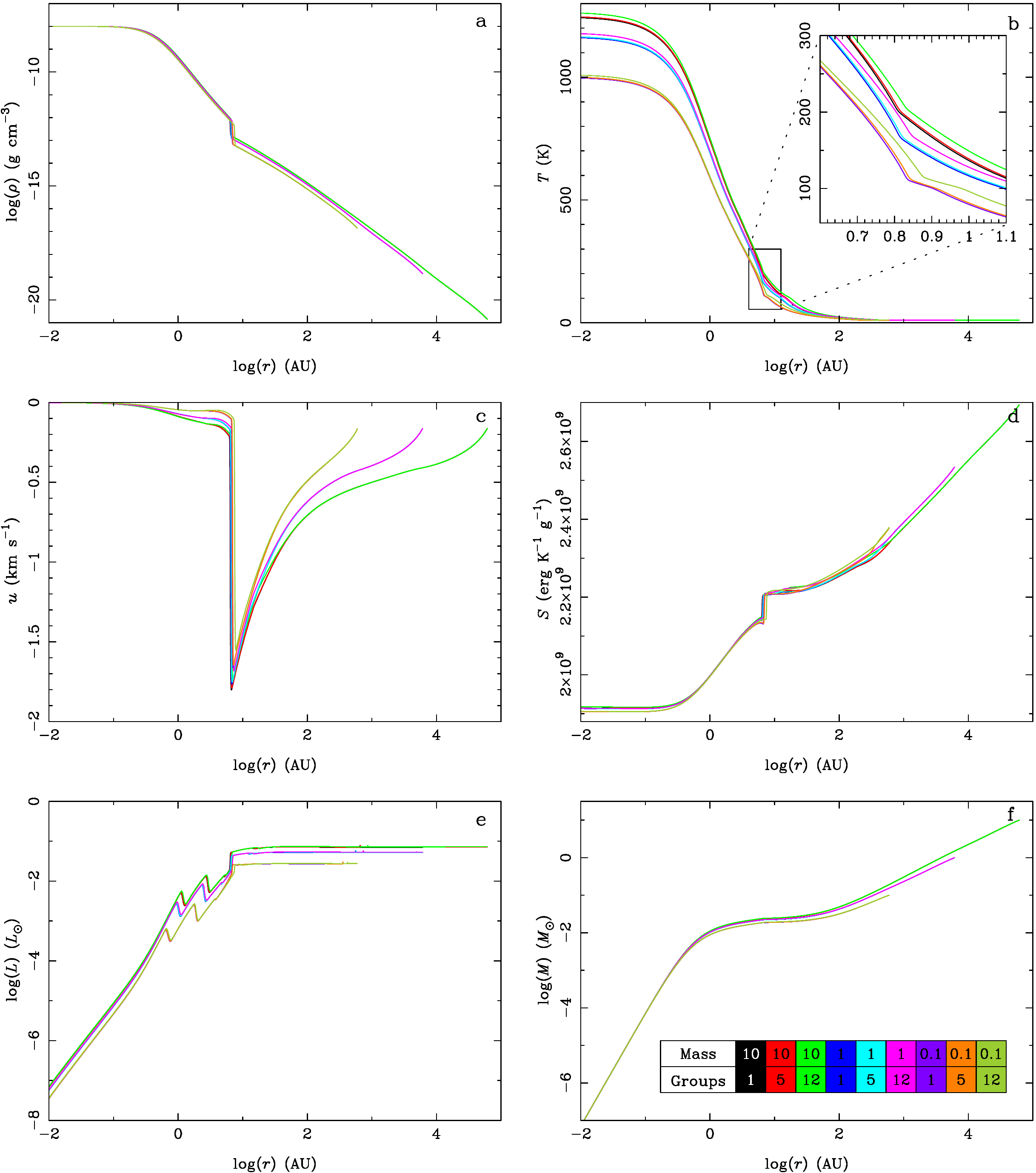}
\caption[First core radial profiles for 3 different cloud masses]{Comparison of the radial profiles during the collapse of a $0.1~\msun$, 1~\msun and 10~\msun dense clouds for a core central density $\rho_{c} = 10^{-8}~\text{g~cm}^{-3}$. From top left to bottom right: (a) density, (b) gas temperature, (c) velocity, (d) entropy, (e) radiative flux and (f) enclosed mass. The legend in the (f) panel explains the curve colour-coding: the top row is the mass of the initial cloud and the bottom row is the number of frequency groups used in each simulation.}
\label{fig:collapse_all_masses}
\end{figure*}

In order to check the universality of the results above, we also performed simulations of the collapse of a 0.1~\msun and a 10~\msun cloud. As mentioned in section~\ref{sec:init_cond}, the initial setup is identical to the 1~\msun case, except for the system's initial radius. The same opacities and frequency groups were also used. The results for the simulations using 1, 5 and 12 frequency groups are shown in Fig.~\ref{fig:collapse_all_masses}, along with the previous results from the 1~\msun case.

The first aspect to notice is that the results are strikingly similar to the 1~\msun case, for both 0.1~\msun and a 10~\msun clouds. The properties of the first core for the new cloud masses are listed in Table~\ref{tab:core-props}, and this illustrates the point further. The shock radii and masses of the first core are very similar (differences of $\sim 20\%$). Some trends do appear in the results; for all the core properties, except the entropy, we note some, although weak, dependence upon the mass of the parent cloud, this effect being the most obvious for the core's central temperature (see Fig.~\ref{fig:collapse_all_masses}b). Figure~\ref{fig:collapse_all_masses}e also shows the more massive parent clouds producing slightly more luminous cores, as was noted by \citet{masunaga1998}.

The second trend in the results listed in Table~\ref{tab:core-props} is that the 12-group simulations always appear to yield hotter, larger and more massive cores compared to the grey runs, but once again those differences remain very small. Overall, the properties of the first core, and especially the entropy inside the core, appear to be independent of the size and mass of the initial cloud \citep[cf.][]{masunaga1998}. There are here again no major differences between the grey and multigroup simulations for all the runs.

\section{Conclusions}\label{sec:conclusions}

We have performed multigroup RHD simulations of the gravitational collapse of a 1~\msun cold dense cloud core up to the formation of the first Larson core, reaching a central density of $\rho_{c} = 10^{-8}~\text{g~cm}^{-3}$. Five groups were first used to sample the opacities in the frequency domain and the results were compared to a grey simulation. Only small differences were found between the two runs, with no major structural or evolutionary changes. The main properties of the resulting first core formed in the centre of the grid such as its mass, size and entropy were not changed significantly. We then performed a third simulation using 12 frequency groups, which again did not affect the results in any major fashion. The grey simulation was however found to slightly underestimate the first core's radius, mass and central temperature.

We also performed simulations of the collapse of a 0.1 and 10~\msun parent cloud, in order to confirm the robustness of the results stated above. The properties of the first core were found to be quasi-independent the initial conditions in the cloud; all cores have a radius of $\sim7$ AU, a mass of $\sim2 \times 10^{-2}~\msun$ and an entropy of $\sim 2 \times 10^{9}~\text{erg~K}^{-1}~\text{g}^{-1}$ at their centre. As for the dependency on the radiative transfer model used, there were once again no significant differences between the grey and multigroup simulations, in both 0.1 and 10~\msun cases.

Nevertheless, we did note that the more frequency groups used, the larger the (small) differences were, and that the multigroup simulations always appeared to yield hotter (differences of 1\%), larger (differences of $\sim10$\%) and more massive (differences of 3\%) cores compared to the grey runs. The multigroup simulations also revealed some interesting facts on the distribution of the radiation energy and flux among the different frequency groups. Indeed, the vast majority of the radiative energy and flux inside the first core is contained within the IR frequency group, while the far IR radiation dominates in the outer envelope. This could help identify first cores in IR observations of giant molecular clouds. We wish to point out here that \citet{masunaga1998} were already using frequency-dependent radiative transfer by solving the transfer equation directly, but their method cannot be realistically used for future 3D simulations, unlike our new multigroup scheme.

The grey solution appears to work well in the case of the first stage of collapse since the medium is optically thick for virtually all the frequency groups behind the accretion shock (see Fig.~\ref{fig:collapse12}e) and the gas and radiation temperatures are very closely coupled. Differences are however expected to arise if the gas and radiation are decoupled, and if radiation of very different energies are present at the same time in the system, as is the case for the second phase of the collapse. Indeed, once the temperatures are high enough in the core, the dissociation of molecular hydrogen enables further rapid collapse giving rise to the situation pictured in \citet{stahler1980} where a very hot core produces mainly UV and X-ray radiation which travels through an opacity gap (created from the sublimation of the circumstellar dust at temperatures above 1500 K) and hits the dust front where it gets absorbed (the dust is optically thick to UV radiation) and re-emitted into the IR. This situation can evidently only be realistically modelled with a frequency-dependent treatment of the radiative transfer.

The second phase of collapse requires the use of a more sophisticated equation of state compared to the ideal gas equation and we will make the use of the Saumon-Chabrier-Van-Horn equation of state \citep{saumon1995} in an upcoming paper to model the complicated physics involved. The frequency-dependent radiative transfer will also allow us to study the radiative flux at the accetion shock in great detail and confirm or invalidate the predictions on the inward/outward flux budget of \citet{stahler1980} which have important consequences on the rate at which star formation occurs.

\acknowledgements

The research leading to these results has received funding from the European Research Council under the European Community's Seventh Framework Programme (FP7/2007-2013 Grant Agreement no. 247060). BC greatfully acknowledges support from his postdoctoral fellowship at the Max-Planck-Institut f\"{u}r Astronomie and the support of the CNES postdoctoral fellowship program. The authors would like to thank the referee for useful comments which have helped improved this paper.

\bibliographystyle{aa}

\end{document}